\documentclass{article}[16pt] 
\usepackage{graphicx}
\usepackage{bm}
\usepackage{amsmath,amssymb}
\usepackage{geometry}
\usepackage{times}
\usepackage{setspace}
\providecommand{\keywords}[1]{\textbf{\textit{Keywords--}} #1}
\usepackage{subfigure}
\usepackage{epsfig}

\begin{document}

\title{Determining physical properties of the cell cortex }  %

\author{
A. Saha, M. Nishikawa, M. Behrndt , C.-P. Heisenberg, F. J\"{u}licher, S. W. Grill
}


%
%
%
%
%
%
%
%
\maketitle



\newcommand{\nwc}{\newcommand}
\nwc{\vs}{\vspace}
\nwc{\hs}{\hspace}
\nwc{\la}{\langle}
\nwc{\ra}{\rangle}
\nwc{\lw}{\linewidth}
\nwc{\nn}{\nonumber}
\nwc{\VET}{\tau_{\rm M}}

\nwc{\pd}[2]{\frac{\partial #1}{\partial #2}}
\nwc{\zprl}[3]{Phys. Rev. Lett. ~{\bf #1},~#2~(#3)}
\nwc{\zpre}[3]{Phys. Rev. E ~{\bf #1},~#2~(#3)}
\nwc{\zpra}[3]{Phys. Rev. A ~{\bf #1},~#2~(#3)}
\nwc{\zjsm}[3]{J. Stat. Mech. ~{\bf #1},~#2~(#3)}
\nwc{\zepjb}[3]{Eur. Phys. J. B ~{\bf #1},~#2~(#3)}
\nwc{\zrmp}[3]{Rev. Mod. Phys. ~{\bf #1},~#2~(#3)}
\nwc{\zepl}[3]{Europhys. Lett. ~{\bf #1},~#2~(#3)}
\nwc{\zjsp}[3]{J. Stat. Phys. ~{\bf #1},~#2~(#3)}
\nwc{\zptps}[3]{Prog. Theor. Phys. Suppl. ~{\bf #1},~#2~(#3)}
\nwc{\zpt}[3]{Physics Today ~{\bf #1},~#2~(#3)}
\nwc{\zap}[3]{Adv. Phys. ~{\bf #1},~#2~(#3)}
\nwc{\zjpcm}[3]{J. Phys. Condens. Matter ~{\bf #1},~#2~(#3)}
\nwc{\zjpa}[3]{J. Phys. A  ~{\bf #1},~#2~(#3)}

\newpage
\section{Abstract}
Actin and myosin assemble into a thin layer of a highly dynamic network underneath the membrane of eukaryotic cells. This network generates the forces that drive cell- and tissue-scale morphogenetic processes. The effective material properties of this active network determine large-scale deformations and other morphogenetic events. For example, the characteristic time of stress relaxation  (the Maxwell time $\VET$) in the actomyosin sets the time scale of large-scale deformation of the cortex. Similarly, the characteristic length of stress propagation (the hydrodynamic length $\lambda$) sets the length scale of slow deformations, and a large hydrodynamic length is a prerequisite for long-ranged cortical flows. 
Here we introduce a method to determine physical parameters of the actomyosin cortical layer {\it in vivo}. For this we investigate the relaxation dynamics of the cortex in response to laser ablation in the one-cell-stage {\it C.~elegans} embryo  and in the gastrulating zebrafish embryo. %
These responses can be interpreted using a coarse grained physical description of the cortex in terms of a two dimensional thin film of an active viscoelastic gel. To determine the Maxwell time $\VET$, the hydrodynamic length $\lambda$ and the ratio of active stress $\zeta\Delta\mu$ and per-area friction $\gamma$, we evaluated the response to laser ablation in two different ways: by quantifying flow and density fields as a function of space and time, and by determining the time evolution of the shape of the ablated region.
Importantly, both methods provide best fit physical parameters that are in close agreement with each other and that 
are similar to previous estimates in the two systems. 
Our method provides an accurate and robust means for measuring physical parameters of the actomyosin cortical layer. It can be useful for investigations of actomyosin mechanics at the cellular-scale, but also for providing insights in the active mechanics processes that govern tissue-scale morphogenesis. 
\section{Introduction}

Cells need to adopt their shape in order to drive tissue scale morphogenetic processes \cite{lecuit2007cell,eaton2011cell,hutson2003forces,solon2009pulsed}. A classical example is convergent extension, where cells reshape and intercalate in order to allow an epithelium to shrink in one direction, while expanding in the other \cite{tada2012convergent}. The actomyosin cortex also endows cells with the ability to reshape themselves \cite{salbreux2012}. This thin layer beneath the membrane largely consists of cross-linked actin filaments and non-muscle myosin motor proteins. Importantly, this thin structure can generate active stresses and contract \cite{salbreux2009hydrodynamics, behrndt2012forces, mayer2010}. Active stresses emerge from the force-generation of myosin motors interacting with actin filaments, fuelled by ATP hydrolysis \cite{levayer2012biomechanical}. 
Such active molecular processes lead to the buildup of mechanical stress (${\boldsymbol \sigma^{\rm a}}$) 
on larger scales \cite{koenderink2009active}.  To gain a physical insight into the stresses and the force balances that govern large-scale deformation and flows of the cell cortex, coarse grained continuum mechanical descriptions have played an important role \cite{behrndt2012forces, mayer2010, bergert2015force}. Furthermore, cortical laser ablation (COLA) has emerged as a useful tool for investigating forces and stresses in the cortical layer \cite{hutson2003forces, behrndt2012forces, mayer2010}. The aim of this manuscript is to extend the analysis of the response of the actomyosin cortex to COLA by use of thin film active viscoelastic gel theory \cite{kruse2004asters, kruse2005, callan2011hydrodynamics}, in order to determine physical parameters that characterize the actomyosin cortical layer.

One  biological example where a continuum mechanics description in the framework of active gels has been useful is epiboly in zebrafish gastrulation  \cite{behrndt2012forces}. Here, a ring of actomyosin cortex  forms on the surface of the yolk cell \cite{koppen2006coordinated}. This ring contracts to generate circumferential stresses, but also to drive flow of actomyosin into the ring. The mechanical stresses that are generated in this process play an important role to pull the connected enveloping layer of cells (EVL) of the blastomere from the animal pole of the embryo towards the vegetal one. 
The stress distribution and the emerging large-scale actomyosin flow velocity field can be understood using thin active gel theory. Cortical tension can also be investigated in experimental terms by cortical laser ablation (COLA). Anisotropies of recoil velocities can be related to tension anisotropies resulting from flow, and viscous tension and the emerging flow fields depend on the large-scale physical parameters of the actomyosin cortical layer. Therefore, in order to understand the mechanics that underlie flow and deformation of the actomysoin cortical layer it is key to determine its large scale physical properties, which has been difficult to achieve. 

A second example that highlights the role of actomyosin in morphogenesis is the polarizing one-cell stage {\it C.~elegans} embryo \cite{cuenca2003polarization, cheeks2004c}. Here, gradients of active tension generate cortical flows that lead to the establishment of anterior-posterior cell polarity, which in turn is key for the subsequent asymmetric cell division \cite{munro2004cortical, goehring2011polarization}. Notably, cortical flow in this system can be well described by
a thin film active gel theory  \cite{mayer2010}. Also here, COLA has permitted the characterization of tension profiles \cite{mayer2010}. However, an estimation of other physical parameters that characterize the actomyosin cortex, such as the effective 2D viscosity $\eta$, the friction coefficient  $\gamma$ with respect to  the membrane and the cytosol have remained elusive. Furthermore, the cortex behaves as an elastic solid on short times, while it is essentially viscous on long times \cite{humphrey2002active}. Thus,  a dynamic description needs to take into account that the actomyosin cortical layer is viscoelastic. A simple model for viscoelastic behaviors which is limited to the slowest relaxation processes
is the Maxwell model which can be incorporated in the theory of active gels \cite{kruse2004asters, kruse2005, callan2011hydrodynamics}. 
Viscoelastic behavior of the cortex stems from the continuous remodeling of the cortical network by the turnover of actin filaments and other cortical constituents \cite{bausch1998local, wen2011polymer, broedersz2010cross}.
This turnover occurs under the influence of regulatory and signaling molecules \cite{levayer2012biomechanical} . Because of turnover, the strain energy stored in the cortex elastic stress relaxes and the corresponding strain energy is dissipated \cite{tawada1991protein, bormuth2009protein}.  The corresponding time scale of stress relaxation (the Maxwell time $\VET$) is an important physical quantity that determines cortex behavior. Mechanical perturbations 
that persist on times large compared to $\VET$ will lead to viscous deformations and flows \cite{mayer2010}. If, however a mechanical perturbation persist only for times that are shorter than $\VET$,  strain energy will be stored and the the cortical layer will respond elastically. To conclude, the time scale of cortical remodelling determines the characteristic timescale of stress relaxation $\VET$, which in turn governs the cortical response to mechanical perturbations.

An second important parameter characterizing the coarse-grained spatiotemporal dynamics of the actomyosin cortex is its hydrodynamic length, $\lambda = \sqrt{\eta / \gamma}$. This length determines the range of stress propagation and sets the correlation length of the cortical flow field. Hence, a large hydrodynamic length $\lambda$ leads to long ranged cortical flows, which, for the case of {\it C.~elegans}, is important for polarization~\cite{goehring2011polarization}. In this manuscript we sought to determine key coarse grained physical parameters of the cell cortex , by use of COLA experiments and by use of a theoretical description of the cortical layer in terms of a two dimensional active viscoelastic gel. We apply our method to two model systems: the actomyosin ring that drives zebrafish epiboly, and the actomyosin cortical layer in the single cell embryo of {\it C.~elegans} that drives cell polarization. We measure distinct physical properties of the cortex in the two systems, thus demonstrating a broad applicability of our method. 

\section{Materials and Methods}

\subsection{{\it C.~elegans} strain and sample preparation}
To image NMY-2 in single cell embryos, we used transgenic line LP133 ({\it nmy-2(cp8[NMY-2::GFP + unc-119(+)])I; unc-119 (ed3) III)}. {\it C.~elegans} maintenance and handling was as previously described \cite{brenner1974genetics}. We cultured the line at 20 C$^{\circ}$ and shifted temperature up to 24 C$^{\circ}$ 24 hours prior to microscope imaging and COLA. Embryos were dissected in M9 buffer (0.1 M NaCl and 4 \% sucrose) and mounted onto the agar pads (2 \% agarose in water) to squish the embryos gently. All experiments were performed at 23 - 24 C$^{\circ}$. 

\subsection{Zebrafish transgenic lines and sample preparation}
To visualize NMY-2 in the yolk syntial layer (YSL) of zebrafish embryos throughout epiboly, we used  transgenic line Tg(actb2:myl12.1-EGFP) \cite{maitre2012adhesion}. Maintenance of the fish line as well as embryo collection were conducted as previously described \cite{westerfield2007zebrafish}. Zebrafish embryos were incubated at 25-31$^{\circ}$C in E3 medium and staged according to morphological criteria \cite{kimmel1995stages}. For imaging and the performance of COLA embryos were mounted in 1 \% low melting point agarose (Invitrogen) inside E3 medium on a glass bottom petri dish (MatTek).

\subsection{Imaging and COLA}
Zebrafish embryos and {\it C.~elegans} embryos were imaged and ablated using modified versions of previously described spinning disk laser ablation systems \cite{behrndt2012forces, mayer2010}. In brief, the spinning disc system (Andor Revolution Imaging System, Yokogawa CSU-X1) was assembled onto a the Axio Observer Z1 inverted microscope (Zeiss) equipped with an 63x water immersion objective. Fluorescent images were acquired by an EMCCD (Andor iXon) at the specified time intervals (for {\it C.~elegans}: 1 sec, for zebrafish 0.5 sec). The pulsed 355nm-UV laser (Powerchip, Teem Photonics) with a repetition rate of 1kHz was coupled into the Axio Observer and steered for point-wise ablation by galvanometric mirrors (Lightning DS, Cambridge Technology).  A custom-built LabView program integrated all devices for simultaneous COLA and imaging. To see the cortical response in {\it C.~elegans} embryos, we applied 20 pulses for each points spaced every 0.5 $\mu m$ along a 10 $\mu m$  line.  The cut lines were chosen to be parallel with the long axis of the embryos, the future AP axis. For zebrafish embryos we applied 25 pulses per point spaced at 0.5 $\mu m$ along a 20 $\mu m$ line. The cut line was placed within the YSL actomyosin ring at a distance of 20 $\mu m$ from the EVL margin and parallel to that margin. The intensity of the UV laser was adjusted to achieve successful COLA in zebrafish embryos. Successful COLA was characterized by the visible opening of the cortex in response to the cut, subsequent recovery of the actomyosin cortex within the cut opening and no `wound healing' response, as previously described \cite{behrndt2012forces, mayer2010}. 

\subsection{Comparison to theory}
We convert the NMY-2 fluorescence intensity $I(x, y, t)$ to a scalar height field $h(x, y, t)$ according to 
\begin{equation}
h(x, y, t) = \frac{I(x, y, t) - I_0}{I_{\rm max}(t) - I_0}.
\end{equation}
Here, $I_0$ denotes the `background' signal obtained from the average intensity within the box located at the center of the cut opening in the first post-cut frame, $I_{\rm max}(t)$ is the maximum intensity in each recorded image, and  $x$ and $y$ denoting spatial cartesian coordinates within the plane of the cortical layer and $t$ denoting time. Note that this conversion assumes that myosin intensity is proportional to cortex height.

To determine the best fit non-dimensionalized model parameters $\alpha_1, \alpha_2, \alpha_3$ and the characteristic time $\tau_{\rm a}$ (see Appendix for the detail), we performed the least square fitting for the COLA responses in experimental observation by theoretical one. We evaluated the spatial velocity profile at a time just following the COLA, temporal evolution of the width of the cut opening boundary, and the recovery timecourse (see Fig.~\ref{fig2}). Fitting was performed iteratively by using the Nelder-Mead method \cite{nelder1965simplex}. 
First we obtained the best fit value of $\tau_{\rm a}$ with the arbitrary values of $\alpha's$ by fitting the recovery timecourse. We then nondimensionalized the experimental time course with determined $\tau_{\rm a}$ value and obtained the best fit values of $\alpha's$ through fitting the spatial velocity profile and the time evolution of the cut opening boundary. In next iteration step, we determined the best fit value of $\tau_{\rm a}$ with $\alpha's$ values obtained in the previous iteration step. We stopped the iteration after the convergence of the fitting parameters. 

In parallel to this, we tested to find the best fit parameter values by comparing the temporal development of the cut boundary between the experimental data and theoretical prediction. We determined the cut opening boundary by automatically detecting the cut opening region. The edge points were detected by using active contour method, a built-in function in Matlab (Mathworks). Then we obtained the best fit parameter values iteratively by using the Nelder-Mead method \cite{nelder1965simplex}. In each iterations, we first determined the best fit value of $\tau_{\rm a}$ by fitting the recovery curve with the arbitrary values of $\alpha's$, to set the time scale for the non-dimensionalization. Then we computed the cut response numerically with the given parameter values. We next compared the edge points of the cut opening region between the experimental observation, $(x_i^{\rm e}(n), y_i^{\rm e}(n))$, and theoretical prediction, $(x_j^{\rm t}(n), y_j^{\rm t}(n))$, at the given time point, n. 
We computed the pairwise distances between $(x_i^{\rm e}(n), y_i^{\rm e}(n))$ and $(x_j^{\rm t}(n), y_j^{\rm t}(n))$, $d_{ij}(n)$, and then calculated the minimum distances, with respect to each j, $d_i^{\rm m}(n) = {\rm min}_{j} [d_{ij}(n)]$. Then we took sum with respect to $i$, and average for the frames in analysis, to get the distance measure,  
\begin{equation}
D = \frac{1}{N} \sum_n \sum_i d_i^{\rm m}(n).
\end{equation}
where N represents the number of frames in analysis. We minimized $D$ to obtain the best fit parameter values of $\alpha$'s with the given value of $\tau_{\rm a}$. The iteration is stopped after the convergence of the parameter values. 

\section{Results}
\subsection{Cortical response to COLA}

We first sought to quantify precisely the spatiotemporal dynamics of cortical non-muscle myosin II (NMY-2) that arises in response to ablating the cortex along a line. For this we used spinning disc microscopy to image NMY-2 fluorescence in combination with a UV laser ablation setup to sever the cortical layer along a line \cite{behrndt2012forces, mayer2010, colombelli2004ultraviolet}. We recorded the spatiotemporal evolution of the surrounding NMY-2 in the cortex to following the resealing process. 

\begin{figure}[h]
\includegraphics[width=15.5cm,height=9cm]{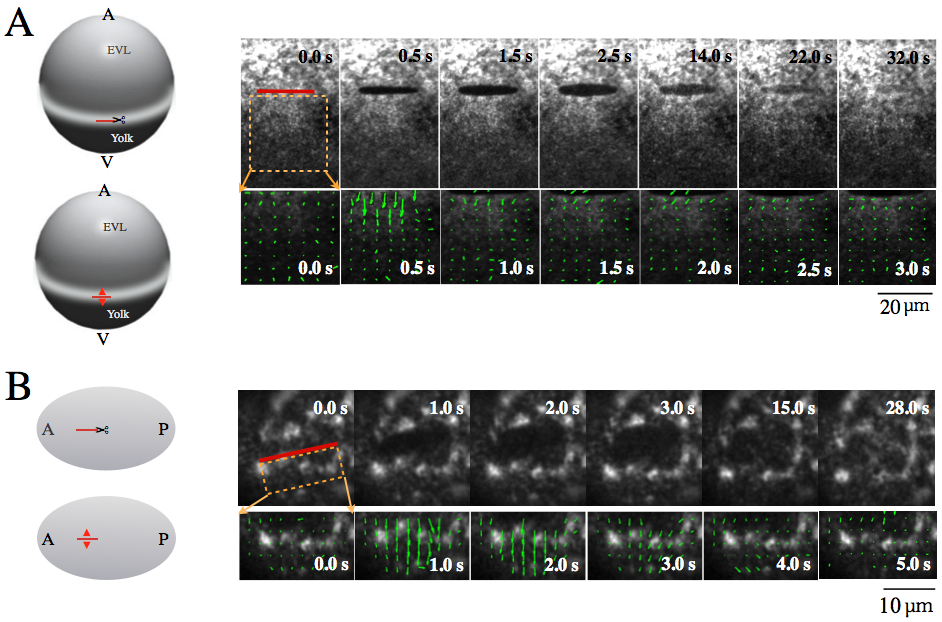}
\caption{{\bf Probing cortical tension in the actomyosin ring of gastrulating zebrafish ({\bf {A}}) and in the {\it {C.~elegans}} ({\bf {B}})actomyosin  cortex of single cell embryos by COLA.} {\bf {A}} Left, schematic of COLA (scissors) in the actomyosin ring of zebrafish. COLA is performed along a 20 $\mu m$ line (red line) at the stage of $65$\% epiboly. Red arrow-heads, direction of cortical recoil after ablation. Right, upper images show a time series of cortical NMY-2-GFP following ablation, lower images show the corresponding velocity fields as determined by PIV. {\bf B} Left, schematic of COLA on the actomyosin cortex the {\it {C.~elegans}} zygote. COLA is performed along a 10 $\mu m$ line in parallel to AP axis of the embryo. Red arrow-heads, direction of cortical recoil after ablation. Right, upper images show a time series of cortical NMY-2-GFP following ablation, lower images show the corresponding velocity fields as determined by PIV.}
\label{fig1}
\end{figure}

In case of the zebrafish embryo, at the stage of $60\%$ - $70\%$ epiboly,  COLA was performed within the YSL actomyosin ring and along a line parallel to EVL.  In case of the {\it C.~elegans} one cell embryo, the cortex was ablated just before the onset of  cortical flow in the anterior half of the embryo and in a direction along AP axis. In both systems the area surrounding the cut was imaged until the hole was no longer visible due to turnover and regrowth (Fig.~\ref{fig1} A and B right upper).
\\
COLA severs all connections within the cortex along the cut line and sets tension in the direction orthogonal to the cut line to zero. This results in tension gradients that drive an outward movement of the adjacent cortex away from the cut line~(Fig.~\ref{fig1} A and B right). Notably, despite significant differences in cortical structure and dynamics,  both  zebrafish and {\it C.~elegans} share large similarities in the overall response to COLA and the spatiotemporal dynamics of cut opening and resealing. In both systems the outward movement of the cut boundary lasts for several seconds and turns the cut line into an approximately elliptically shaped opening. Furthermore, the adjacent cortex moves outward with a velocity that decays over time, followed by cortical regrowth within the cleared region until no visible mark of the COLA procedure remains. 

\begin{figure}[h]
\begin{center}
\includegraphics[width=10.0cm,height=10.0cm]{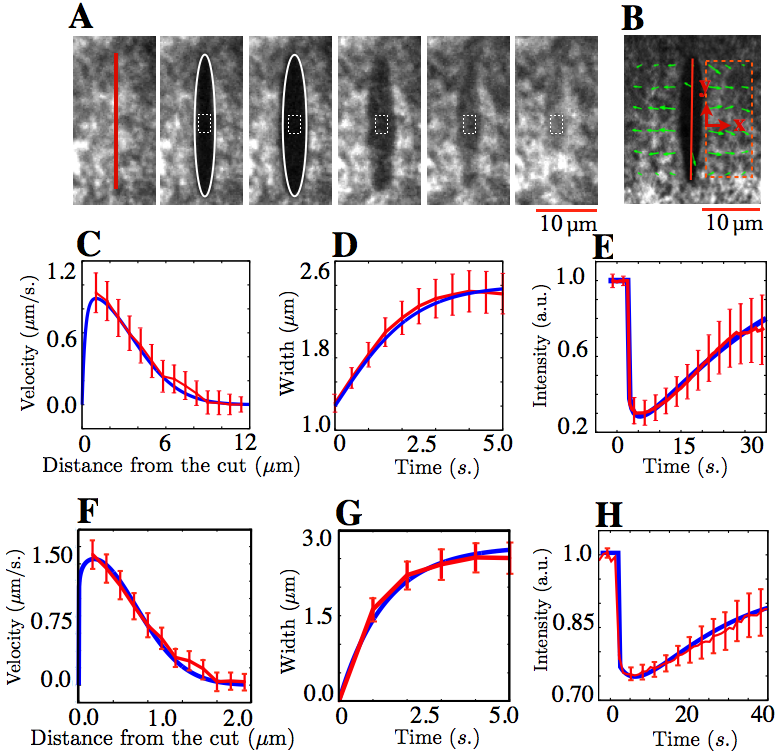}
\caption{{\bf {COLA response in zebrafish and {\bf {{\it C. elegans}}} :}} {\bf A} and {\bf B}, Illustration of the analysis of the cortical response to COLA. {\bf A}, the time evolution of the cut boundary opening was characterized by fitting an ellipse to the cut boundary and measuring the minor axis of the fitted ellipse, white solid line. To analyze the regrowth of the cortical NMY-2 after the cut we measured the average fluorescence intensity within the box indicated by the white broken line. The size of the box is $1.8\times1.4~\mu m$ for zebrafish and $1.4\times1.2~\mu m$ for {\it C.~elegans}. {\bf {B}} To obtain the velocity profile along the direction perpendicular to the cut line (red line), the x components of the velocity vectors inside the rectangular area surrounded by the orange broken line were averaged in the y direction (box size, zebrafish: $27\times 13~\mu m$ , {\it {C. elegans}}: $12 \times 10~ \mu m$). {\bf C} and {\bf F}, velocity profile along the direction perpendicular to the cut line at the time just after the cut for zebrafish ({\bf C}) and {\it C. elegans} ({\bf F}).
{\bf D} and {\bf G}, time evolution of the minor axis of the fitted ellipse to the cut boundary for zebrafish ({\bf C}) and {\it C. elegans} ({\bf F}).
{\bf E}  and {\bf H}, average concentration of  cortical NMY-2 as a function of time after the cut opening for zebrafish ({\bf C}) and {\it C. elegans} ({\bf F}). 
Red,  experimental results in zebrafish ({\bf {C-E}} ) and {\it C.~elegans} ({\bf F-H}). Error bars represents standard error with n=15 for all cases. Blue lines, theory curves utilizing the obtained least square fit parameters, see main text. }
\label{fig2}
\end{center}
\end{figure}

To extract the characteristics of the response of the cortex to COLA and the characteristics of the resealing process, we analyzed the spatiotemporal dynamics of cortical NMY-2 after COLA in  several ways. We first determined the outward velocity of cortical NMY-2 adjacent to the cut line by Particle Image Velocimetry (PIV) at a time just following the cut, as indicated by the arrows in Fig.~\ref{fig1} A and B right lower panels. Remarkably, the velocity profile along the direction perpendicular to the cut line was not uniform, but rather decayed over a characteristic distance away from the cutline as shown in Fig.~\ref{fig2} C and F. This spatial decay entails information about the characteristic distance $\lambda$ over which mechanical stresses communicate in the 2D  cortical layer. Second, we quantified the temporal evolution of the cut opening. For this, we determined the extent of the opening generated by COLA, by fitting an ellipse to this opening and determining the time evolution the minor radius of this fitted ellipse as a measure of the width of the cut opening (Fig.~\ref{fig2}~A). Notably, the width of the cut opening increased with time and reached maximum after approximately 3-5 seconds (Fig.~\ref{fig2}~D and G). The minor radius of the opening grows with a characteristic time governed by processes of stress relaxation in the actomyosin cortical layer. Third, we analyzed NMY-2 regrowth within the opening, by quantifying the average fluorescence intensity in a box that is placed at the center of the COLA opening (white broken line in Fig. \ref{fig2} A). Following an initial drop in intensity due to ablation and outward movement, NMY-2 intensity gradually recovered over approximately 30 seconds (Fig. \ref{fig2}~E and H). Similar time scales have been observed in FRAP measurements that characterize NMY-2 turnover \cite{mayer2010}.

These observations lend credence to the assumption that the cortex behaves as an active viscoelastic material. The shape evolution of the cut opening is determined largely by elastic properties of the cortex as well as active tension provided by NMY-2. The velocity decay away from the cut line is largely determined by the decay length of tension in the layer, and thus by $\gamma$ and $\lambda$ . Finally, re-association of myosin at the cut site is determined by actomyosin turnover. In what follows, we will compare the dynamics of opening and regrowth, as determined in our experiments by characterizing the spatial decay of the velocity field and the time evolution of the cut opening width and the myosin levels at the center of the hole, with theory.

\subsection{Physical description of the actomyosin cortex}

We next sought to calculate the cut response in a coarse grained physical description of 
the cortical layer. Considering that the thickness of the cortex is small compared to the size of the cell we describe the actomyosin cortex as an active 2D viscoelastic compressible fluid. We introduce a scalar field $h(x,y,t)$ which denotes the local height of the cortical layer in the $z$ direction, with $x$ and $y$ denoting spatial cartesian coordinates within the plane of the cortical layer, and $t$ denoting time. Considering a viscoelastic isotropic active fluid in the plane 
and integrating over the height of the cortex we write the following constitutive equation: 
\begin{eqnarray}
&&\left(1+\VET D_t\right)(\sigma_{ij}-\sigma^{\rm a}_{ij})
\label{consti}\\
\nonumber
&=&\eta(\partial_{i}v_{j}+\partial_{j}v_{i}-\partial_{k}v_{k}\delta_{ij})+\eta_{\rm b}\partial_{k}v_{k}\delta_{ij} \, ,
\end{eqnarray}
where indices $i,j,k=x,y$. The dynamic variables are the 2D stress tensor ${\boldsymbol \sigma}$, the 2D active stress tensor ${\boldsymbol \sigma^{\rm a}}$ and the 2D velocity field ${\bf v}$. The material properties are characterized by the 2D shear viscocity $\eta$, 2D bulk viscocity $\eta_{\rm b}$ and a characteristic Maxwell time of stress relaxation $\VET$. Here, $D_t = \partial_t + {v_{i}} \partial_{i}$ denotes the material time derivative. Turnover of the film is captured by the dynamics of the height given by 
\begin{eqnarray}
\frac{\partial h}{\partial t}= - \partial_{i}({v_{i}}h)+\frac{h_0-h}{\tau_{\rm a}}  \, ,
\label{height}  
\end{eqnarray}        
where the first term on the r.h.s. accounts for advection of the gel by cortical flow,  and the second term describes turnover which relaxes to the steady state hight $h_0$. Note that both the turnover  time $\tau_{\rm a}$ and the relaxed height $h_0$ depend on actin and myosin turnover. The force balance equation reads \cite{salbreux2009hydrodynamics, behrndt2012forces, mayer2010, bois2011, kumar2014pulsatory}
\begin{eqnarray}
\partial_{i}\sigma_{ij}=\gamma v_{j} \, ,
\label{force}
\end{eqnarray} 
where inertial forces have been neglected and $\gamma$ is a friction coefficient that describes frictional interactions between the cortex and it its surrounding cytosol and membrane.

We consider an active gel that is incompressible in 3D, with constant density through the height of the cortical layer. In this case, both the active stress $\sigma^{\rm a}$ and the 2D viscosities $\eta$ and $\eta_{\rm b}$ of the cortex are proportional to cortex height $h$ and are written as 
\begin{eqnarray}
\sigma^{\rm a}_{ij} & = & \frac{\xi\Delta\mu \, h}{h_0}\delta_{ij}\, ,\\
\eta(h) & = & \frac{\eta_{0} h}{h_0}\, ,\\
\eta_{\rm b}(h) & = & \frac{\eta_{{\rm b}0} h}{h_0}\, .
\end{eqnarray}

Here, $\xi\Delta\mu$ denotes the isotropic active stress generated through ATP consumption of myosin, positive for contraction and dependent on the change in chemical potential associated with ATP hydrolyosis $\Delta\mu$. Furthermore, $\eta_0$ and $\eta_{{\rm b}0}$ denote the shear and bulk viscosities of the layer when $h=h_0$. 3D incompressibility condition couples divergences in the 2D flow velocity field $\bf v$ to change the cortex height $h$ according to Eq.~\ref{height}, and we set $\eta_{0}/\eta_{{\rm b}0}=3$. Eqs.~\ref{consti}-\ref{force} complete the model.  For nondimensionalization we choose the characteristic time of regrowth  $\tau_{\rm a}$ and the COLA cut length $\lambda_{\rm c}$ as the respective time and length scales. Model equations with dimensionless variables can be rewritten with dimensionless parameters,  $\alpha_1 = \frac{\lambda}{\lambda_{\rm c}}$, $\alpha_2=\frac{\VET}{\tau_{\rm a}}$, $\alpha_3=\frac{\xi\Delta\mu\tau_{\rm a}}{\gamma\lambda_{\rm c}^2}$ (see Appendix).

\begin{figure}[h]
\includegraphics[width=15cm,height=6.43cm]{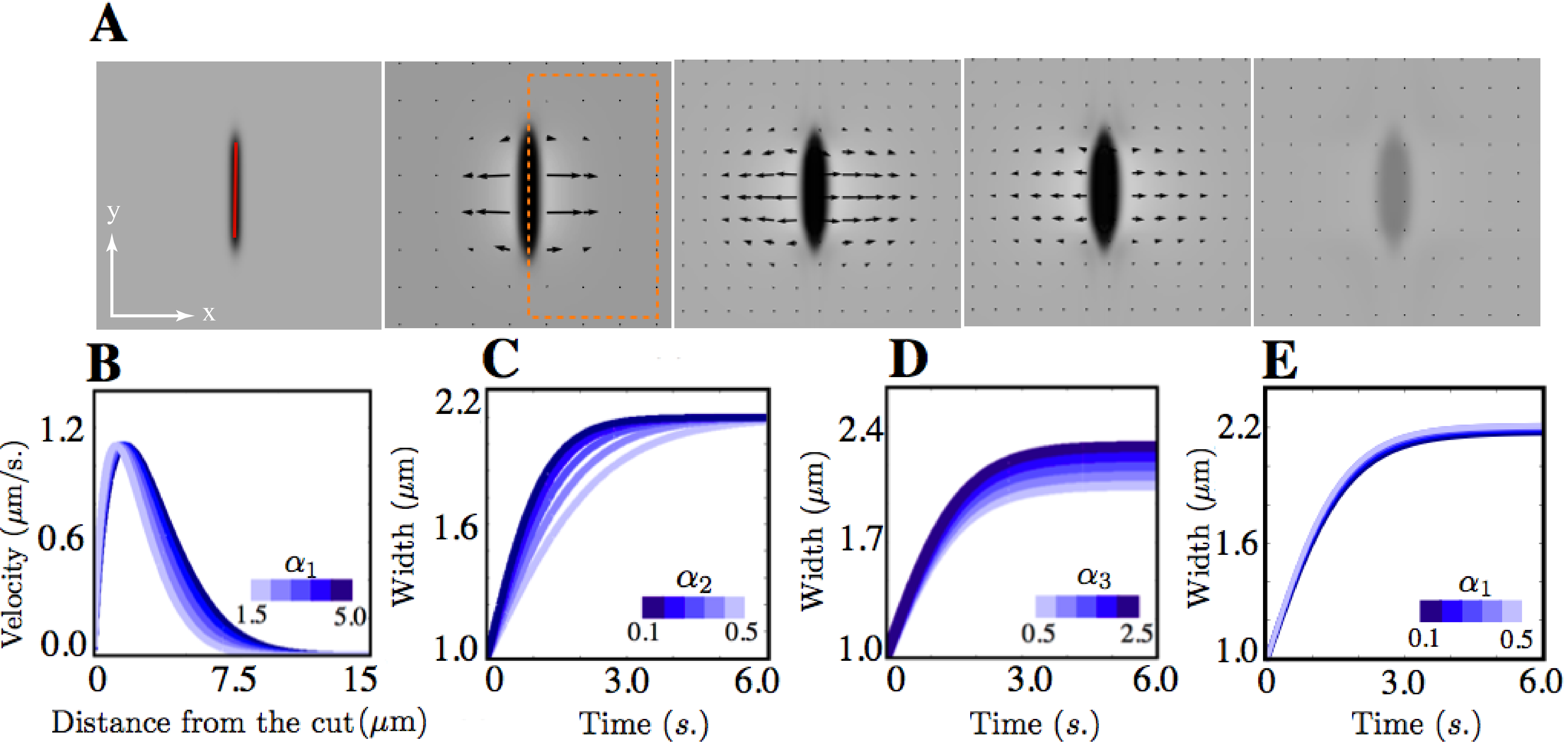}
\caption{{\bf {Numerical calculation of the spatiotemporal response of a 2D active viscoelastic fluid in response to COLA:}} {\bf {(A)}} Representative calculated nondimensional height (grey) and velocity field (arrows) of the cortex at times $0, 0.02, 0.04, 0.06, 0.08, $ and $0.68$ following COLA by setting the height to zero in the dark region around the red line in the leftmost image (see main text). {\bf {(B)}}, Effect of variation of $\alpha_1$ on the spatial velocity profile at  $0.5 s$ after cut. The $x$ component of velocity vectors was averaged along the $y$ direction and within the box indicated by the broken orange line in the second panel of {\bf {(A)}}. {\bf {(C-E)}}, Effect of variation of $\alpha_2$ and $\alpha_3$ on the time evolution of the half-width of the cut opening, determined by fitting an ellipse to the opening boundary. Note that changing $\alpha_1$ has a small  effect on time evolution of the width of the opening ({\bf {E}}). Unless otherwise specified parameter values are $\alpha_1=2.0$, $\alpha_2=0.25$, $\alpha_3=4.5$. For {\bf {(B-E)}} we have dimensionalized time by $\tau_{\rm a}=25 s$ and length by $\lambda_c = 25 \mu m$.  }
\label{fig3}
\end{figure}

 We next asked whether this description can reproduce the relaxation dynamics in response to COLA that we observed in our experiments. To this end we numerically solved non-dimensional versions of Eq. \ref{consti}-\ref{force} (see Appendix) in a rectangular box of width $L$ ( $\sim 15 \lambda_{\rm c}$) in $x$ and $y$, with periodic boundary conditions. As an initial condition we choose uniform height and stress fields that correspond to the unperturbed stationary solution with ${\bf v} = 0$. To account for COLA, this homogenous initial condition is  perturbed at $t=0$ by  setting height $h$ to zero within a thin rectangular strip of length $\lambda_{\rm c}$ and width $\sim 0.1\lambda_{\rm c}$. We then computed the resulting spatiotemporal dynamics. Fig.~\ref{fig3} A displays the time evolution of the resultant height and velocity fields. We observe that 1) the velocity field is not uniform through the cortex but decayed over a distance from the cut line; 2) the width of the boundary initially grows and reaches a maximum; and 3) $h(x,y,t)$ in the the cut region recovers to steady state values on long times.  These observations are consistent with those that we have made in our experiments, which lends credence to our approach.

Next we analyzed our results from theory in terms of the spatial profile of the velocity in the direction perpendicular to the cut line, the growth of the cut boundary, and the recovery of the cut region in a manner that is similar to how we analyzed the experimental data. Fig.~\ref{fig3} B to E demonstrates that the essential features observed in the corresponding graphs from our experiments (Fig.~\ref{fig2} C to H) are reproduced in our theory.  We next asked how changes of physical parameters of the cortical layer impact on the spatial profile of the velocity in the direction perpendicular to the cut line, the growth of the cut boundary, and the recovery of cortex height. To this end we performed numerical simulations of COLA  with different values of $\alpha's$. We find that increasing $\lambda$ leads to a corresponding increase of the spatial decay length of the velocity profile in a direction perpendicular to the cut line (Fig.~\ref{fig3} B), but has little impact on the growth timecourse of cut boundary~(Fig.~\ref{fig3} E). Furthermore, increasing $\VET$ results in a corresponding increase in the relaxation time to reach the maximal width of the opening (Fig.~\ref{fig3} C), with little effect on the final size of the hole. Taken together, key aspects of the relaxation process after COLA are separately determined by the characteristic length and time, $\lambda$ and $\VET$ of the cortical layer.

\subsection{Comparison of theory and experiment for determination of physical parameters} 

\begin{figure}[h]
\includegraphics[width=15.5cm,height=9cm]{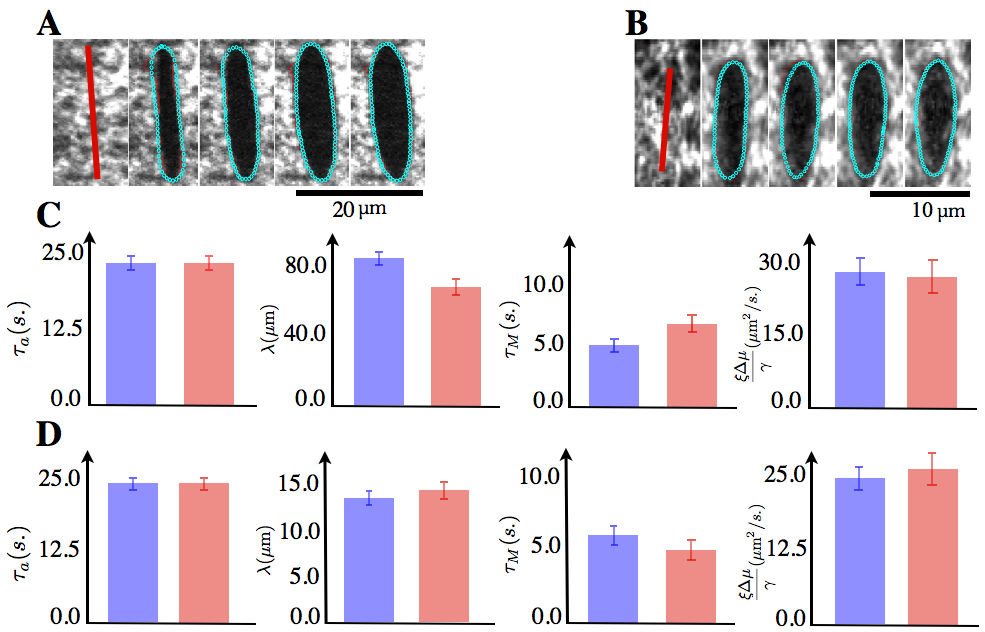}
\caption{{\bf {Determination of physical parameters by comparing the COLA responses observed in experiment to the computed response from the theory:}} {\bf A} and {\bf B,} 
Representative examples of fitting the cut opening boundary in response to COLA of the actomyosin ring in zebrafish during epiboly ({\bf A}) and the cortex in single cell embryo of {\it C.~elegans} ({\bf B}). Red points, automatically detected boundaries of the cut opening. Distances between the computed and the detected boundaries were minimized to find the best fit parameter values. Cyan, the theoretical boundaries that best fits the experimental ones. Images are one second apart. {\bf C} and {\bf D,} Comparison of physical parameter values between the two methods utilized. In red, the best fit parameter values were determined by comparing the experimentally determined regrowth of the cut region, spatial decay of the outward velocity profile, and the time evolution of the cut opening boundary to theoretical prediction (see Figure 2) . In blue, the best fit parameter values obtained by comparing the cut opening boundaries between experiment and theory, see ({\bf A}) and ({\bf B}). Note that both fitting procedures converge to similar values. Error bars are the standard error of the mean with $n=15$ for zebrafish and 10 for {\it C.~elegans}.}
\label{fig4}
\end{figure}

Next we determine coarse grained physical parameters of the cell cortex both in {\it C.~elegans} single cell embryo and zebrafish actomyosin ring, by comparing the experimentally determined cortical response to COLA in terms of regrowth of the cut region, spatial decay of the outward velocity profile, and the time evolution of the cut opening boundary to theory.  We note that the height variable $h(x,y,t)$ in the theory is proportional to the per-area-concentration of molecules that are uniformly distributed along the $z$ direction.  Since the thickness of the cortical layer is smaller than the focal depth of the confocal microscope (approximately $1 \mu m$), the  fluorescence intensity of cortical NMY-2, $I(x,y, t)$ in our microscope images is proportional to the 2D concentration of  NMY-2, integrated over the height of the cortical layer. Hence the local 2D concentration $I(x,y, t)$ of NMY-2 in our microscope images is proportional to $h(x,y,t)$. We determined the best fit non-dimensionalized model parameters $\alpha_1= \frac{\lambda}{\lambda_{\rm c}}, \alpha_2=\frac{\VET}{\tau_{\rm a}} , \alpha_3=\frac{\xi\Delta\mu\tau_{a}}{\gamma\lambda_{\rm c}^2}$ within a nonlinear least-square fitting scheme \cite{nelder1965simplex} by iteratively calculating the theoretical responses that best fit the experimental profiles (see blue curves in Fig. \ref{fig3} C to H in comparison to the experimental profile given by red curves). In Fig.~\ref{fig4} C and D, we report the physical parameters of the cell cortex ($\tau_{\rm a}, \lambda, \VET, \frac{\xi\Delta\mu}{\gamma}$) from the best fit model parameters. We obtained $\tau_{\rm a} = 23.4 \pm 1.63 s, \lambda = 67.6 \pm 8.2 \mu m, \VET = 5.6 \pm 1.2 s$ and $\frac{\xi\Delta\mu}{\gamma} = 25.9 \pm 6.4 \mu m^2/s$ for zebrafish, and $\tau_{\rm a} = 24.2 \pm 1.36 s, \lambda = 14.4 \pm 1.5 \mu m, \VET = 4.48 \pm 1.25 s$ and $\frac{\xi\Delta\mu}{\gamma} = 25.42 \pm 4.4 \mu m^2/s$ for {\it C. elegans}. Note that these values were obtained by fitting each individual experiment (n=15 for both systems), and we report the respective averages $\pm$ standard error. Note also that the hydrodynamic lengths of $\sim 68 \mu m$ for the zebrafish actomyosin ring and $\sim 14 \mu m$ for the {\it C.~elegans} cortex as well as the turnover times of $\sim 25 s$ show a close agreement with previous investigations \cite{behrndt2012forces, mayer2010}. This supports our method of extracting values of physical parameters by use of fitting the measured COLA response to that expected from theory. 

The good agreement between theory and experiment indicates that the shape evolution of the cut opening might entail a sufficient amount of information for accurately determining both  $\lambda$ and $\VET$. Since the spatial profile of the velocity field is governed by hydrodynamic length, $\lambda$, the temporal change of the shape of the cut opening boundary is likely to be affected by $\lambda$. On the other hand, the temporal decay of the outward velocity is characterized by the time scale of the stress relaxation, $\VET$.  Thus it is possible that the shape evolution of the cut boundary is to a large extent governed by these two physical parameters. In agreement with this statement, the shapes of the cut opening boundary are distinct between the zebrafish actomyosin ring and {\it C.~elegans} embryo, which might reflect the respective differences in physical parameters. 

Therefore we asked if it is possible to determine the hydrodynamic length $\lambda$ and the time scale of stress relaxation $\VET$ from the shape evolution of the cut opening boundary. To this end we determined the $\alpha$'s by comparing the spatiotemporal development of the cut opening boundary shapes in experiment and theory. We detected the shape of the cut opening boundary observed in experiment to evaluate the difference with the shape from the theoretical prediction at the given parameter values. We computed the minimum distance from $i$th edge point detected in the experiment to the edge points obtained from the theory at the $n$th time frame, $d_i^{\rm m}(n)$, (see Materials and Methods for details). We then evaluated a merit function, $D = \frac{1}{N} \sum_n \sum_i d_i^{\rm m}(n)$, where $N$ is the number frames analysed (a useful choice for $N$ was between  8 and 12 for a  zebrafish, and between 4 and 6 for {\it C.~elegans}).  $D$ was minimized iteratively to find the best fit values of $\alpha$'s and $\tau_{\rm a}$. Fig~\ref{fig4} A and B shows the shape of the cut opening boundary from the theory with the best fit parameter values with respect to the minimization of $D$, successfully reproducing the shape evolution of the cut opening boundary observed in experiment. Consequently we obtained non-dimensionalized model parameters $\alpha_1, \alpha_2 , \alpha_3$, as well as the physical parameters $\tau_{\rm a} = 23.4 \pm 1.66 s, \lambda = 80.2 \pm 6.8 \mu m, \VET = 4.95 \pm 1.0 s$ and $\frac{\xi\Delta\mu}{\gamma} = 26.89 \pm 5.2 \mu m^2/s$ for zebrafish and $\tau_{\rm a} = 24.2 \pm 1.33 s, \lambda = 13.6 \pm 1.3 \mu m, \VET = 5.37 \pm 1.2 s$ and $\frac{\xi\Delta\mu}{\gamma} = 24.65 \pm 3.7 \mu m^2/s$ for {\it C.~elegans}. Again, these values were obtained by fitting each individual experiment (n=15 for zebrafish and n= 10 for {\it C. elegans}, in the latter not all of our experimental datasets converged in the fitting procedure), and we report the respective averages $\pm$ standard error. The values of the parameters of the actomyosin network are in close agreement between this method of determination, and the method used before, see Fig.~\ref{fig4} C and D. We conclude that the shape evolution of the cut opening boundary entails a sufficient amount of information to determine the entire set of non-dimensional physical parameters, and provides a second means of determining physical parameters of the actomyosin cortical layer.

\section{Discussion}
While COLA is a powerful tool to characterize the tension in the actomyosin cortex, obtaining the physical parameters to describe the cortical mechanics relies on an appropriate analytical method. Here we present a strategy to determine the physical parameters of actomyosin cortex from the cortical response to COLA. The method relies on the fitting of the COLA response to the theoretical prediction from the 2D active viscoelastic fluid model. We show that the best fit value of the hydrodynamic length, $\lambda$'s are in close agreement with previous estimates both for zebrafish actomyosin ring and {\it C.~elegans} single cell embryo \cite{mayer2010, behrndt2012forces}. Notably we can determine a whole set of parameters in a single experiment, in contrast the previous estimates that require the ensemble averaging for the flow profile, and myosin distribution. As shown in Fig.~\ref{fig4}, the standard error of the best fit parameters are small in both  methods, signifying the overall robustness of the approach. In addition, our method does not require any assumptions for boundary conditions in the flow and myosin profiles. Taken together, our method allows the precise determination of the physical parameters. 

Importantly, our method also allows to determine the characteristic time of stress relaxation, $\VET$, which governs the time scale between elastic and viscous regime. $\VET$ sets the timescale for the large-scale movement of the actomyosin cortex, and morphogenetic deformations of the cortex that are driven by actives stresses in the layer generally occurs on  time scales larger than $\VET$. Recently active microrheology has been performed to measure the storage and loss moduli by manipulating a bead injected inside the cell \cite{guo2014probing}. This method allows the precise determination of the characteristic time of stress relaxation of the cytoplasm. However a typical diameter of bead used in the active microrheology measurement, $\sim 500nm$ is bigger than the thickness of the cortical layer, one cannot distinguish the cortical and cytoplasmic viscoelastic properties in the measurement. In contrast, our method allows us to directly determine  $\VET$ of the cortex. 

In summary, we have developed a robust and accurate method with broad applicability to determine the physical parameters of cortical mechanics from COLA experiment in conjunction with the coarse grained mechanical theory. It provides us with the large-scale and biologically relevant parameters in terms of morphogenetic mechanics. Given the simple description of the cortex used in the paper, we suggest that our method can be applied for complex multicellular systems such as epithelial tissues to determine the physical parameters that describes the tissue mechanics. 

\section{Appendix: Non-dimensional equations}
We rescaled the time and the spatial coordinate by setting the cut length ($\lambda_{\rm c}$) as length scale and $\tau_{\rm a}$ as time scale of the system.  
Eqs.\ref{consti}-\ref{force} can be rewritten as, 
\begin{eqnarray}
&&\left(1+\alpha_2 D_t\right)(\sigma_{ij} -
\sigma^a\delta_{ij}) \label{consti-nond} = \alpha_1^{2}h(\partial_{i}\partial_{k}\sigma_{kj}+\partial_{j}\partial_{k}\sigma_{ki}+2\partial_{k}\partial_{m}\sigma_{mk}\delta_{ij})\\
&&\frac{\partial h}{\partial t}=-\alpha_3\partial_{j}\left(h{\partial_i \sigma_{ij}}\right)+1-h\\
&&v_{j}=\alpha_3\partial_{i}\sigma_{ij}
\end{eqnarray} 
where $\lambda=\sqrt{\frac{\eta_0}{\gamma}}$ represents the hydrodynamic length. 
$\alpha_1=\frac{\lambda}{\lambda_{\rm c}}$, $\alpha_2=\frac{\VET}{\tau_{\rm a}}$ and  $\alpha_3=\frac{\xi\Delta\mu\tau_{\rm a}}{\gamma\lambda_{\rm c}^2}$ are three independent, dimensionless parameters of the model.

\section{Acknowledgments}
We are grateful to Daniel Dickinson for providing LP133 {\it C.~elegans} strain. We thank G. Salbreux, V. K. Krishnamurthy and J. S. Bois for fruitful discussions. SWG acknowledges support by grant no. 281903 from the European Research Council (ERC) and by grant GR~7271/2-1 from the Deutsche Forschungsgemeinschaft (DFG). SWG and CPH acknowledge support through a grant from the Fonds zur F\"orderung der wissenschaftlichen Forschung (FWF) and the DFG (I930-B20).

\section{Author Contributions}
AS, FJ and SWG designed the research, MB and MN performed experiments under supervision of CPH and SWG, AS and MN developed the theory and performed numerical simulations as well as the comparison to data under supervision of FJ and SWG, MN and SWG wrote the paper with the help of AS and with support from FJ.

\bibliographystyle{unsrt}
\bibliography{AS_paper1}

\end{document}